\providecommand{\doi}[1]{doi: #1}

\providecommand{\doi}[1]{doi: #1}

\documentclass[a4paper,11pt]{article}
\usepackage{aaskaiid}
\usepackage{orcidlink}

\title{Cosmic Magnetism Science with the SKA }
\ShortTitle{Cosmic Magnetism Science with the SKA}

\author[1]{Tessa Vernstrom\orcidlink{0000-0001-7093-3875}}
\author[2,3]{Jennifer L. West\orcidlink{0000-0001-7722-8458}}
\author[4]{Cathy Horellou\orcidlink{orcid:0000-0002-3533-8584}}

\affiliation[1]{Australia Telescope National Facility, CSIRO, Space and Astronomy, PO Box 1130, Bentley, WA 6102, Australia}
\emailAdd{tessa.vernstrom@csiro.au}
\affiliation[2]{Dominion Radio Astrophysical Observatory, Herzberg Astronomy \& Astrophysics, National Research Council Canada, P.O. Box 248, Penticton, BC V2A 6J9, Canada}
\affiliation[3]{
School of Natural Sciences, University of Tasmania, PO Box 807, Sandy Bay, TAS 7006, Australia}
\emailAdd{jennifer.west@utas.edu.au}
\affiliation[4]{Chalmers University of Technology, Department of Physics and Astronomy, Division of Astronomy and Plasma Physics, Onsala Space Observatory, SE-439 92 Onsala, Sweden}
\emailAdd{cathy.horellou@chalmers.se}
\abstract{Magnetic fields are a fundamental component of astrophysical systems, yet many key questions about their origin, amplification, and role in structure formation and evolution remain unresolved. The SKA will mark a transformational step forward in addressing these questions, enabling studies of cosmic magnetism across a large range of spatial scales and environments. This overview summarizes the main science cases in the Cosmic Magnetism Science Working Group, which cover a huge breadth of scales from the smallest scales governing planet and star formation, all the way up to the large-scale structure of the Universe. The chapter summarizes the main observational techniques for studying magnetic fields, including direct polarization imaging, Faraday rotation, rotation-measure grids, and Zeeman splitting. We also address fundamental considerations of these studies including SKA-Low vs SKA-Mid and wide-area vs deep observing strategies.  }

\begin{document}
\maketitle

\section{Introduction}




SKA will be transformative for studies of cosmic magnetism. Its unprecedented combination of sensitivity, wide frequency coverage, angular resolution, and 
polarimetric capabilities will enable the detection and characterization of magnetic fields across an enormous range of spatial scales and environments. In particular, the SKA will enable dense rotation measure (RM) grids, high-fidelity Faraday tomography, and the detection of faint diffuse polarized emission, opening new discovery space for understanding the origin and evolution of cosmic magnetic fields.

Faraday rotation 
manifests itself as a change in polarization orientation as a function of frequency (or more precisely as wavelength squared). This phenomenon provides 
a powerful probe 
of line-of-sight magnetic fields, with other measurements, such as synchrotron brightness and Zeeman splitting, providing means to deduce 3D magnetic field structures. Broad wavelength coverage, especially at low frequencies, provides opportunities for Faraday depth tomography, separating Faraday-rotating structures along the line of sight, which is critical to providing sensitivity and resolution in Faraday depth space. High spatial resolution mitigates issues caused by mixing of this 
complex-valued polarization 
within the telescope beam, greatly improving the ability to disentangle the complicated information.

The contributions in this volume reflect the breadth of science enabled by the SKA for cosmic magnetism. These studies span magnetic fields in galaxy clusters and the large-scale structure of the Universe, individual galaxies and active galactic nuclei, and the Milky Way, as well as new observational techniques and analysis methods. Together, these contributions demonstrate the complementary approaches required to build a comprehensive understanding of cosmic magnetism.

In this chapter, we provide an overview of the contributions of the Cosmic Magnetism Science Working Group to \textit{Advancing Astrophysics with the SKA II}. We summarize the key science goals, highlight connections between the contributions, and discuss the observational capabilities required to fully realize the potential of the SKA for cosmic magnetism science.



\section{Main Science Cases}

Here we provide a summary of the contributions to this volume, exploring magnetism in galaxy clusters and large-scale structure, individual galaxies and active galactic nuclei, and the Milky Way, as well as the observational techniques required to study these systems. The  diverse science cases in this volume are unified by the common goal of understanding the origin, evolution, and role of magnetic fields across cosmic time and the techniques used to probe them \citep[see also][]{2020Galax...8...53H}.

\subsection{Galaxy Clusters and Large-scale Structure}
Galaxy clusters and the large-scale structure of the Universe provide unique laboratories for studying magnetic fields on megaparsec scales. These environments probe the amplification of magnetic fields through turbulence, mergers, and large-scale motions, and provide insights into the origin of cosmic magnetism.

\citet{Vacca01.2026.SKA.halos} 
investigate the detectability of polarized emission in radio halos and filaments in the era of SKA observations. Using simulations of merging galaxy clusters and filamentary structures, they explore the potential for the SKA to reveal polarized emission from the intracluster medium and the surrounding cosmic web. These studies highlight the importance of sensitivity and angular resolution for detecting faint polarized structures in low surface brightness environments.

\citet{AritraBasu01.2026.SKA}  examine the role of fluctuation dynamos in amplifying magnetic fields in the intracluster medium. Their work explores the expected polarized synchrotron emission arising from turbulent magnetic fields and evaluates the probability of detecting these signals with the SKA. These studies provide key predictions for observational signatures of turbulence-driven magnetic field amplification.

The impact of dense RM grids on the study of galaxy clusters and groups is explored by \citet{Loi01.2026.SKA}. By leveraging large samples of polarized background sources, RM grids enable statistical studies of magnetic fields in clusters and groups, as well as their evolution with cosmic time. These measurements complement direct observations of diffuse emission and provide an independent probe of magnetized large-scale structures.

\citet{Kurahara01.2026.SKA} present a multi-wavelength approach to studying turbulence and magnetic fields in galaxy clusters, combining radio observations with X-ray measurements from XRISM. This work highlights the synergy between radio polarimetry and X-ray observations in constraining the turbulent energy budget and magnetic field properties of the intracluster medium.

\citet{OSullivan01.2026.SKA} explore the capabilities of SKA-Low to construct dense RM grids, focusing on low-frequency polarized source counts and their implications for constraining the origin of cosmic magnetism. Low-frequency observations offer enhanced sensitivity to Faraday rotation, enabling precise measurements of weak magnetic fields in large-scale structures such as filaments of the cosmic web.

\citet{Akahori01.2026.SKA.clusters} 
investigate magnetic reconnection in galaxy clusters as a potential mechanism for particle acceleration and energy dissipation. Polarization observations provide a powerful probe of magnetic reconnection processes, and SKA observations will offer new insights into the microphysics of cluster magnetic fields.

\citet{Vacca02.2026.SKA.bayesian} 
further explore the application of Bayesian statistical techniques to polarization surveys, enabling statistical studies of magnetization in the large-scale structure of the Universe. These approaches leverage large samples of polarized sources to probe weak magnetic fields beyond the detection limits of individual objects.


\subsection{Galaxies and Active Galactic Nuclei}
Magnetic fields play an important role in galaxy evolution, influencing star formation, gas dynamics, and feedback processes. Several contributions in this volume focus on magnetic fields in individual galaxies and active galactic nuclei.

\citet{Sawada-Satoh01.2026.SKA} examine the detection of weakly polarized emission from magnetized circumnuclear matter in active galactic nuclei. These observations will probe magnetic fields in the immediate vicinity of supermassive black holes, offering insights into accretion and jet-launching mechanisms.

\citet{Sakemi01.2026.SKA} investigates the detectability of polarized emission associated with ultraluminous X-ray sources and intermediate-mass black holes. These observations provide a unique opportunity to study magnetized jets and bubbles in these extreme environments.

The use of strong gravitational lensing systems to probe magnetic fields in distant galaxies, providing insights into the evolution of magnetism over cosmic time is explored in \citet{Mao01.2026.SKA}.

\citet{Mao01.2026.SKA} also propose a comprehensive polarimetric survey of nearby galaxies, combining diffuse polarized emission and RM grid measurements to map magnetic fields within galactic disks and halos. SKA will also provide the opportunity of measurements of Zeeman splitting in the Local Group of galaxies, 
which is advantageous since Galactic studies are mostly confined to the plane \citep[see][]{Robishaw01.2026.SKA}. 
Nearby galaxies provide 
a holistic, external view of spiral galaxy magnetic fields 
(vs the interior view that we have in the Milky Way) while still providing access to angular scales that allow investigations of smaller, turbulent scales. The high angular resolution of SKA will enable investigations of small-scale galactic magnetic structures, which is presented in \citet{Ma01.2026.SKA}.

\subsection{Milky Way Studies}

In contrast to the external view afforded by nearby galaxies, our location within a spiral arm of the Milky Way provides the unique opportunity for up-close studies of magnetic fields at much smaller scales starting from sub-pc details of star forming regions. Several chapters explore the possibilities of probing small scales in the Milky Way, including 
those of 
\citet{Sun01.2026.SKA.GMF} and \citet{Ma01.2026.SKA}, which present the importance of understanding the small-scale galactic magnetic fields and the ability of SKA to detect them. \cite{Bracco01.2026.SKA} addresses the prospects of direct detection of synchrotron emission from dense starless cores in molecular clouds. \citet{Tahani01.2026.SKA} present general techniques that involve exploiting the unique properties of polarized radiation and the relationships with other observable tracers to overcome projection effects in order to uncover 3D magnetic field geometry in nearby Galactic structures including cores, filaments, and bubbles. \citet{Sun02.2026.SKA.LocalBubble} specifically explore the rich possibilities of probing the 3D structure of the Local Bubble using SKA-Low. 

The potential of using SKA to measure Zeeman splitting in the Milky Way to provide direct estimates of the magnetic field strength and direction in the interstellar medium is described in 
\cite{Robishaw01.2026.SKA} and \cite{Bourke01.2026.SKA}.

Finally, all-sky RM-grids have historically been used to probe the geometry of the Galactic Milky Way magnetic field at the largest scale. \cite{Sun01.2026.SKA.GMF} discuss the improvements that a higher density RM-grid from SKA will allow. This is particularly true when used in combination with images of the diffuse polarized emission both from SKA and also from single-dish telescopes to complete the measurements, where interferometers miss flux at the largest angular scales.

\section{Techniques}

The study of cosmic magnetism relies on a diverse set of observational techniques, each probing different aspects of magnetic field strength, structure, and evolution. The unprecedented sensitivity, frequency coverage, and polarimetric capabilities of the SKA telescopes will require significant advancements to these techniques. Several contributions in this volume focus on the development and application of these methods.

\subsection{Faraday Tomography}

Faraday tomography has emerged as one of the most powerful tools for studying cosmic magnetism. Broadband polarimetric observations enable the reconstruction of Faraday depth spectra, allowing multiple magnetized structures along the line of sight to be separated and characterized. This technique is particularly important for disentangling complex environments such as the interstellar medium, galaxy clusters, and large-scale structure.

\citet{Carcamo01.2026.SKA} provide an overview of Faraday tomography techniques in the SKA era, highlighting the exciting potential of broadband polarimetric observations. The SKA's wide frequency coverage and sensitivity will enable high-resolution Faraday depth measurements, allowing detailed studies of magnetic field structure and turbulence across a wide range of astrophysical environments.

\subsection{Direct detection of diffuse polarized emission}
Direct imaging of diffuse polarized emission provides information that is complementary to RM grid studies. Using observations of the intensity and angle of polarized synchrotron emission, and the associated polarized fractions (highlighting depolarization effects), reveal magnetic field morphology and turbulence within extended structures such as galaxies, clusters, and filaments.

Several contributions explore the detection of diffuse polarized emission in galaxy clusters and large-scale structure \citep[e.g.][]{Vacca01.2026.SKA.halos, AritraBasu01.2026.SKA}. These studies emphasize the importance of sensitivity to low surface brightness emission and high angular resolution to mitigate beam depolarization. The SKA's capabilities will enable the detection of faint polarized emission from environments that have remained inaccessible to current instruments.

The main source of diffuse polarized emission in the sky is from the interstellar medium of the Milky Way. This emission forms a veil that is not confined to the Galactic plane, but which is truly located in every direction in the sky including at high Galactic latitudes. Increasingly there is an appreciation that these Galactic foregrounds confound studies of extragalactic structures, particularly (but not exclusively) studies of structures at extended angular scales. Several chapters in this volume propose studies that make use of the diffuse emission of the Milky Way \citep[e.g.,][]{Ma01.2026.SKA, Tahani01.2026.SKA, Sun01.2026.SKA.GMF, Sun02.2026.SKA.LocalBubble}, which will not only contribute knowledge of the Milky Way's magnetic field, but also inform studies of 
magnetic fields
across the universe. 

\subsection{RM grids}

RM grids have become a cornerstone of cosmic magnetism studies. By measuring the Faraday rotation of large samples of polarized background sources, RM grids provide a statistical probe of foreground magnetic fields. The dramatic increase in polarized source density expected with the SKA will enable high-fidelity mapping of magnetic fields across the sky.

Measuring the field structure in the foreground via an RM
grid requires background sources with one-component Faraday spectra (i.e., Faraday-simple). Recent surveys demonstrate that the number of Faraday-simple sources decreases with increasing spatial resolution and Faraday resolution. High Faraday resolution is also needed to measure the width of the Faraday components that could originate in the foreground if the angular size of a background source is large. Hence, high spatial resolution \textbf{and} high Faraday resolution are crucial for successful RM grid projects.

Several contributions in this volume highlight the power of RM grid studies. \citet{Sun01.2026.SKA.GMF} presents the case for using RM grids to investigate the large-scale Galactic magnetic field. \citet{OSullivan01.2026.SKA} explore the potential of SKA-Low for constructing dense RM grids at low frequencies, and \citet{Loi01.2026.SKA} examine the use of RM grids to probe magnetic fields in galaxy clusters and groups. These studies demonstrate the importance of large-area surveys and high source densities for probing magnetized structures across cosmic scales. \citet{AritraBasu01.2026.SKA} and \citet{Kurahara01.2026.SKA} also propose specific RM-grid survey setups and predictions. \citet{Carcamo01.2026.SKA} has a dedicated section on the RM-grid construction and SKA projections. From the discussions in these chapters, we provide some estimates of source densities in Table~\ref{tab:rmgrid} and general densities vs sensitivities are shown in Figure~\ref{fig:rmdensity}.

\subsection{Zeeman splitting}
Zeeman splitting provides a direct measurement of magnetic field strength and direction in magnetized gas. While this technique is limited to specific environments, it provides an important complement to Faraday rotation and synchrotron emission studies.

\citet{Robishaw01.2026.SKA} discuss the use of Zeeman splitting with the SKA to measure magnetic fields in the interstellar medium of the Milky Way and external galaxies. The sensitivity of the SKA will enable measurements of weaker magnetic fields and extend Zeeman studies to a wider range of astrophysical environments. A dedicated discussion of Zeeman splitting in Galactic star-forming regions is presented in \citet{Bourke01.2026.SKA}.

\subsection{Modelling and Statistical Approaches}

In addition to observational techniques, modelling and statistical approaches play an important role in interpreting polarization observations. \citet{Chan01.2026.SKA} present a cosmological polarized radiative transfer framework that bridges theory and observations, enabling direct comparisons between models and polarization measurements. Such approaches will become increasingly important as the SKA produces large polarization datasets.


\section{SKA-Low and SKA-Mid}

Traditionally many polarization studies have focused on frequencies around 1~GHz or higher, but low frequencies studies (e.g., with LOw-Frequency ARray, LOFAR) are proving to be extremely valuable for magnetism studies \citep[see][for details of previous work]{OSullivan01.2026.SKA}. At low frequencies, there is a much greater degree of Faraday rotation, which can cause depolarization, thereby reducing the polarized signal (i.e., low polarized fractions). For this reason, higher frequencies were favoured. However, low frequencies offer much higher precision in RM-measurements, which is a game changer particularly when undertaking Faraday tomography, the process through which we can separate the distances where the emission originates. In addition, at higher resolutions, this problem is mitigated (due to less mixing within the beam). The SKA-Low RM Grid may help identify sky regions with minimal contamination by small-scale Galactic magnetic fields, which could be excellent sky windows for future extragalactic magnetism studies \citep{Ma01.2026.SKA}. Several chapters address the truly transformative promise of SKA-Low for magnetism science \citep[e.g.,][]{OSullivan01.2026.SKA, Sun02.2026.SKA.LocalBubble}.

The contributions in this volume make use of a broad range of SKA-Mid frequency 
bands, reflecting the complementary science enabled by different observing 
frequencies. Most RM-grid and broadband polarimetric studies focus on SKA-Mid 
Band~2 (0.95--1.76 GHz), which provides an optimal balance between sensitivity, 
sky coverage, angular resolution, and Faraday depolarization \citep[see ][]{Loi01.2026.SKA,Ma01.2026.SKA,Sakemi01.2026.SKA, Vacca01.2026.SKA.halos,Vacca02.2026.SKA.bayesian}. 
Several studies 
also emphasize the importance of lower-frequency observations in Band~1 
(350--1050 MHz), particularly for Faraday tomography, studies of diffuse 
polarized emission, and probing weak magnetic fields in large-scale structure \citep[see ][]{Akahori01.2026.SKA.clusters, Carcamo01.2026.SKA}. 
Higher frequency observations in Bands~5a (4.6--8.5 GHz) and 5b 
(8.3--15.4 GHz) are particularly important for studies of strongly depolarized 
regions, circumnuclear environments in active galactic nuclei, Zeeman splitting, 
and complex Faraday-thick structures \citep[see ][]{Robishaw01.2026.SKA,Mao01.2026.SKA,AritraBasu01.2026.SKA,Carcamo01.2026.SKA, Ma01.2026.SKA}. Together, the broad frequency coverage of 
SKA-Mid enables complementary probes of magnetic fields across a wide range of 
astrophysical environments.

\section{Survey Strategies: Wide versus Deep}

\begin{table*}
\centering
\caption{Representative summary of proposed observation configurations and predicted RM grid densities based on the estimates made in several of the cosmic magnetism science chapters where explicit RM grid estimates are provided \citep{Carcamo01.2026.SKA, OSullivan01.2026.SKA, Loi01.2026.SKA, Mao01.2026.SKA, Sun01.2026.SKA.GMF, Vacca02.2026.SKA.bayesian}}  
\begin{tabular}{ccccccc}
\hline
 Telescope &  Frequency  & $\sigma$ & Time Required & Time Required & RM Density\\
 & & ($\mu$Jy/beam) &  for AA* (hr) & for AA4 (hr) & (deg$^{-2}$) \\
\hline
 SKA-Mid &  0.95--1.76 GHz &  $0.5$ & 11 & 5 & $\sim400$ \\
 
 SKA-Mid & 0.95--1.76 GHz &  $2$ & 0.7 & 0.3 &$\sim100$ \\

 SKA-Low & 100--350 MHz & $10$ & 1.25 & 0.5 & $\sim6$ \\


\hline
\end{tabular}
\label{tab:rmgrid}
\end{table*}

\begin{figure}
    \centering
    \includegraphics[width=1.0\linewidth]{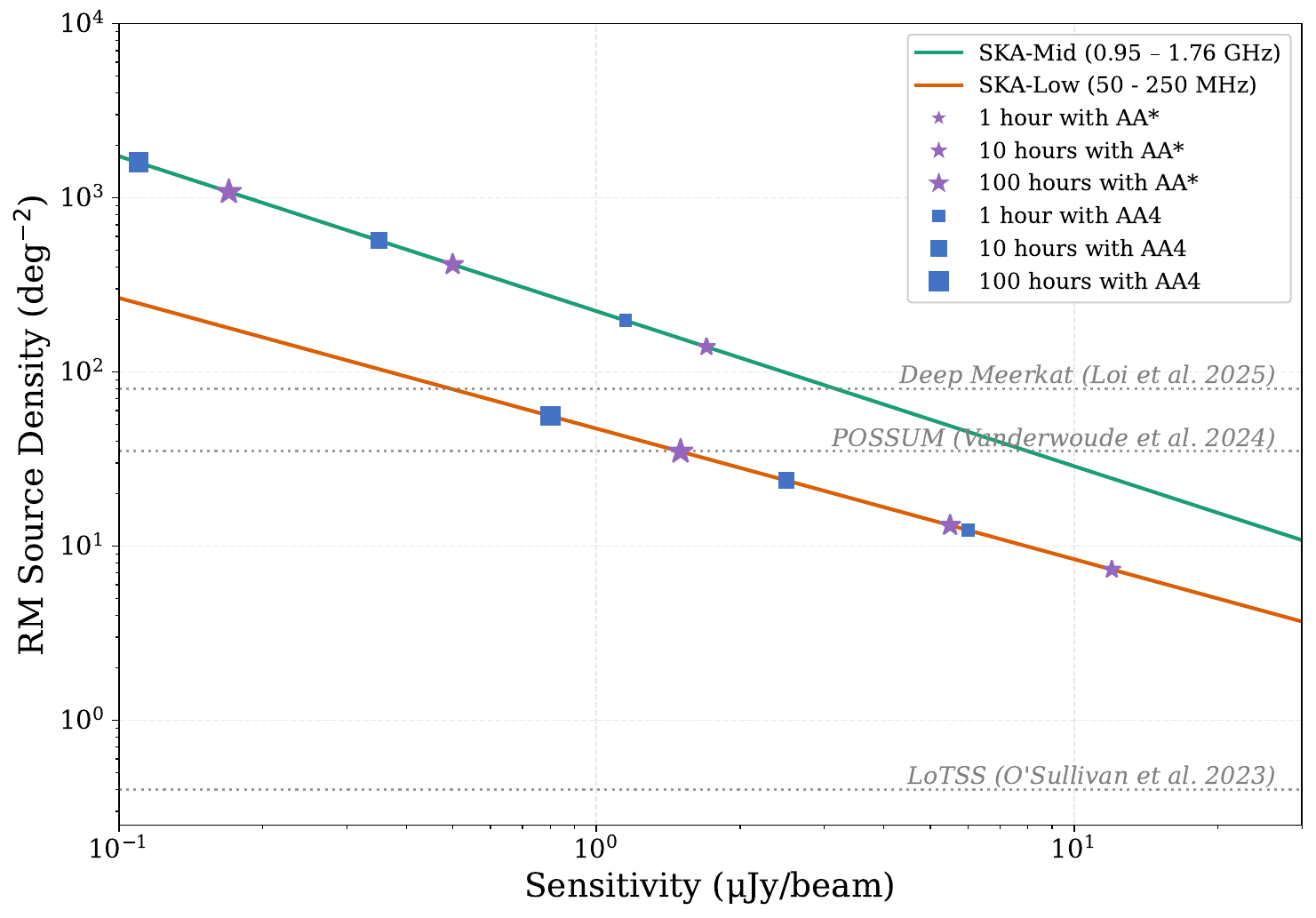}
    \caption{RM source density as a function of sensitivity for SKA-Mid \citep[0.95--1.76~GHz, see][Eq.~5]{Loi01.2026.SKA} and SKA-Low \citep[50--250~MHz, see][Eq.~1]{OSullivan01.2026.SKA} with sensitivities for sample integration times of 1 hour, 10 hours, and 100 hours for AA* and AA4 derived from the SKA sensitivity calculator using robust weighting and $\delta=-40^\circ$. Note that for this plot we use a 5$\sigma$ threshold for both Mid and Low. For comparison, we show representative recent RM source density measurements at $\sim1$~GHz from the Polarisation Sky Survey of the Universe's Magnetism \citep[POSSUM,][]{2024AJ....167..226V, 2025PASA...42...91G} and a deep Meerkat field \citep{2025A&A...694A.125L}; and  at $\sim$150~MHz from the LOFAR Two-metre Sky Survey \citep[LoTSS,][]{2023MNRAS.519.5723O}.}
    \label{fig:rmdensity}
\end{figure}


Survey design plays a critical role in maximizing the scientific return of cosmic magnetism studies. Both wide-area and deep surveys offer complementary approaches for probing magnetic fields across cosmic scales.

Wide-area surveys enable statistical studies of magnetic fields across large volumes of the Universe. Dense RM grids derived from wide surveys allow mapping of magnetic fields in the Milky Way, nearby galaxies, galaxy clusters, and large-scale structure. These surveys are particularly valuable for identifying rare objects, studying large-scale magnetic field structure, and investigating cosmic evolution.

\citet{Sun01.2026.SKA.GMF} advocate for wide-area surveys to study the magnetic field of the Milky Way, emphasizing the importance of uniform sky coverage for reconstructing large-scale magnetic field structure. Similarly, several studies in this volume highlight the importance of wide-area RM grids for probing magnetized structures across cosmic scales.

Deep surveys, on the other hand, provide sensitivity to faint polarized emission and weak magnetic fields. Deep observations are particularly important for detecting diffuse emission in galaxy clusters, filaments, and the intergalactic medium. These observations complement wide-area surveys by probing the low-surface-brightness regime and enabling detailed studies of individual systems. Deep surveys will also deliver denser RM grids, allowing detailed studies of small-scale magnetic structures.

The optimal strategy for cosmic magnetism science will therefore involve a combination of wide and deep surveys. Wide-area surveys will provide statistical samples and large-scale mapping, while deep observations will probe faint structures and detailed magnetic field properties. The flexibility of the SKA will allow both approaches to be pursued, maximizing the scientific return for cosmic magnetism studies.

Several contributions in this volume provide predictions for polarized source densities and resulting RM grid capabilities for different survey strategies including \cite{Carcamo01.2026.SKA, OSullivan01.2026.SKA, Loi01.2026.SKA, Mao01.2026.SKA, Sun01.2026.SKA.GMF, Vacca02.2026.SKA.bayesian}. 
Table~\ref{tab:rmgrid} and Figure~\ref{fig:rmdensity} summarize these predictions. In the first SKA book, \citet{2015aska.confE..92J} predicted that an all sky RM survey with a sensitivity of 4 $\mu$Jy/beam at 2$''$ resolution should provide 7-14 million extragalactic
RMs, which represents a density of approximately 230--460 sources per square degree (assuming 75\% sky coverage). This is a considerably higher than the numbers presented in the present book. Progress in this field since that time \citep[e.g.,][]{2024AJ....167..226V, 2025PASA...42...91G,2025A&A...694A.125L} have shown that the numbers in 2015 were overestimated.

\section{Outlook and Future Directions}

The contributions in this volume highlight the breadth of science that will be enabled by the SKA, from small-scale magnetic fields in the interstellar medium to magnetic fields in galaxy clusters and the large-scale structure of the Universe. These studies demonstrate the importance of combining multiple observational techniques, including Faraday tomography, RM grids, diffuse polarized emission, and Zeeman splitting, to build a comprehensive picture of magnetic fields across cosmic time.

The dramatic increase in sensitivity and survey speed offered by the SKA will enable dense RM grids across large areas of the sky, providing statistical probes of magnetic fields in a wide range of environments. Deep observations will enable the detection of faint polarized emission from diffuse structures, including galaxy clusters, filaments, and the intergalactic medium. These complementary approaches will provide new insights into the origin and evolution of cosmic magnetism.

Broadband observations spanning SKA-Low and SKA-Mid will further enhance these capabilities, enabling high-resolution Faraday tomography and detailed studies of magnetic field structure and turbulence. The combination of wide frequency coverage, high sensitivity, and excellent polarimetric performance will open new discovery space for cosmic magnetism studies.

Studying polarization is uniquely challenging, requiring detailed and intimate knowledge of the instrumental response. Polarized astronomical signals are weak and they must be isolated from much stronger, unpolarized radiation while correcting for complex, direction-dependent instrumental and environmental effects. Misalignments of telescope components, off-axis primary beam deviations, and cross-talk between electronic signals can all cause errors that introduce artificial polarized signals (e.g., ``leakage'') that corrupt the data fidelity \citep[e.g. see][]{Robishaw01.2026.SKA}. These effects must be carefully calibrated, requiring time and effort. The reward is an extremely rich data product that encodes fundamental information about the formation and evolution of the universe, and will improve the data quality for all of the data products from the instrument. 

Together, these advances will address fundamental questions about cosmic magnetism, including the origin of magnetic fields, their amplification through dynamical processes, and their role in galaxy formation and large-scale structure. The SKA will therefore transform our understanding of magnetic fields across the Universe and establish cosmic magnetism as a central component of astrophysical research in the coming decades.

\section*{Acknowledgments} \label{sec:ack}
The authors gratefully acknowledge the magnetism chapter authors and members of the Cosmic Magnetism Science Working Group who proof read this chapter and provided comments on the text. These include Takuya Akahroi, Rainer Beck, Tyler Bourke, Andrea Bracco, Miguel Cárcamo, Francesca Loi, Yik Ki (Jackie) Ma, Ann Mao, Gina Panopoulou, Tim Robishaw, and Lawrence Rudnick.

\bibliographystyle{abbrvnat-maxbibnames4}
\setlength{\bibsep}{0pt plus 0.2ex} 
\bibliography{chapter}

\end{document}